\documentclass[11pt,aps,pra]{revtex4-1}
\usepackage{amsmath}    
\usepackage{graphicx}   

\def\beq{\begin{equation}}
\def\eeq{\end{equation}}

\begin{document}
\title{Shape resonances of Be$^-$ and Mg$^-$ investigated with method of analytic continuation}
\author{Roman \v{C}ur\'{\i}k} \email{roman.curik@jh-inst.cas.cz}
\affiliation{J. Heyrovsk\'{y} Institute of Physical Chemistry, ASCR,
Dolej\v{s}kova 3, 18223 Prague, Czech Republic}
\author{Ivana Paidarov\'{a}} 
\affiliation{J. Heyrovsk\'{y} Institute of Physical Chemistry, ASCR,
Dolej\v{s}kova 3, 18223 Prague, Czech Republic}
\author{Ji\v{r}\'{\i} Hor\'a\v{c}ek} \email{horacek@mbox.troja.mff.cuni.cz}
\affiliation{Institute of Theoretical Physics, Faculty of Mathematics and Physics, 
Charles University in Prague, V Hole\v{s}vi\v{c}k\'{a}ch 2, 180 00 Prague, Czech Republic}

\date{\today}

\begin{abstract}
The regularized method of analytic continuation is used to study the low-energy
negative ion states of beryllium (configuration 2$s^2\varepsilon p\; ^2P$) 
and magnesium (configuration 3$s^2\varepsilon p\; ^2P$) atoms. 
The method applies an additional perturbation potential and it
requires only routine bound-state multi-electron quantum calculations.
Such computations are accessible by most of the free or commercial quantum
chemistry software available for atoms and molecules.
The perturbation potential is implemented as a spherical Gaussian function
with a fixed width. Stability of the analytic continuation technique 
with respect to the width and with respect to the input range of electron affinities 
is studied in detail. The computed resonance
parameters $E_r$=0.282 eV, $\Gamma$=0.316 eV for the $2p$ state of 
Be$^-$ and $E_r$=0.188 eV, $\Gamma$=0.167 for the $3p$ state of Mg$^-$,
agree well with the best results obtained by much more elaborated and 
computationally demanding present-day methods.
\end{abstract}

\maketitle

\section{\label{sec-intro}Introduction}
Resonances in electron-atom or electron-molecule scattering, also addressed as transient
negative ions, have attracted attention over the last decades. It is because these
temporary states provide a pathway for electron-driven chemistry via dissociative
electron attachment (DEA) and therefore, applications can be found in chemistry of
the planetary atmospheres
\cite{Carelli_FAG_2013},
nanolithography in microelectronic device fabrication
\cite{Dorp_PCCP_2012,Thorman_Oddur_BJN_2015},
 and in cancer research where these states provide a mechanism for the DNA damage by 
low-energy electrons
\cite{Sanche_PRL_2003,Sanche_PRL_2006}.

Accurate calculation of energies and lifetimes of the resonances represents a
challenging task that is more complicated than the determination of energies
of the bound atomic or molecular states. Temporary negative ions differ from the bound
states in two important respects: (i) they are not stable and decay into various
continua, (ii) corresponding poles of the $S$-matrix are complex and they are
expressed by
$E=E_r - i\Gamma/2$.
There have been numerous studies published using several methods for determination
of the resonance energies and widths. Stabilization methods
\cite{Taylor_Hazi_PRA_1976,Hazi_Taylor_PRA_1970,Hazi_Kurilla_PRA_1981,Frey_Simons_JCP_1986}
search for a region of stability of the energies with respect to different confining
parameters. Stieltjes imaging technique
\cite{Hazi_Kurilla_PRA_1981} allows to represent the resonant state by a square-integrable
basis and the width is defined by the resonance-continuum coupling. 
Complex rotation methods
\cite{Moiseyev_PR_1998,McCurdy_Rescigno_PRL_1978,Reinhardt_ARPC_1982}
and the methods employing complex absorbing potential
\cite{Riss_Meyer_JPB_1993,Feurbacher_Cederbaum_JCP_2003}
compute complex resonant energy as an eigenvalue of a complex, non-Hermitian
Hamiltonian.

Recently the method of analytic continuation in coupling constant (ACCC) 
\cite{Kukulin_Krasnopolsky_JPA_1977,Krasnopolsky_Kukulin_PL_1978,KKH_book}
has been applied to several molecular targets, such as N$_2$
\cite{Horacek_MU_PRA_2010,White_HGMC_JCP_2017}, ethylene
\cite{JIR_JPCA_2014,Sommerfeld_MHPEM_JCTC_2017},
and amino acids \cite{Papp_Horacek_CP_2013}.
Furthermore, the known low-energy analytic structure of the resonance
was incorporated into the inverse ACCC (IACCC) method providing so-called
regularized analytic continuation (RAC) method. The RAC method was
successfully employed for determination of $\pi^*$ resonances of
acetylene \cite{JIR_EPJD_2016} and diacetylene \cite{JIR_JCP_2015} anions,
proving that the ACCC method can yield accurate resonance energies and
widths for various molecular systems using data obtained with standard
quantum chemistry codes.

Common feature of all methods of analytic continuation is an application of the
perturbation potential $\lambda V$ to the multi-electron Hamiltonian $H$,
i.e. $H \rightarrow H + \lambda V$. 
The role of this attractive perturbation is to transform the resonant
state into a bound state.
Although the RAC method was developed
for strictly short-range perturbation $V$, authors were able to 
successfully use the Coulomb potential in its stead
\cite{JIR_EPJD_2016,JIR_JCP_2015}. This obvious inconsistency can
yield reasonable results, because in practical applications the
perturbation potential is often projected on a finite set of short-range
basis functions, e.g. Gaussian functions used by the quantum chemistry
software. However, so obtained weakly-bound states need to be examined
carefully because they may, in fact, be Rydberg states supported by 
the basis and the long-range tail of the Coulomb perturbation $V$
\cite{JIR_JCP_2015}.
Such states need to be excluded from the continuation procedure as they
do not represent a resonance transferred to a bound state.
In order to avoid such complications, in the present study we adopt
a short-range perturbation potential in a form the Gaussian function
\beq
\label{eq-Gauss1}
V(r) = -\lambda e^{-\alpha r^2}\;.
\eeq
This choice of the perturbation was recently evaluated
by \citeauthor{White_HGMC_JCP_2017} \cite{White_HGMC_JCP_2017}
and applied to the well-known $^2\Pi_g$ resonance of N$_2^-$.
Furthermore, \citeauthor{Sommerfeld_Ehara_JCP_2015} 
\cite{Sommerfeld_Ehara_JCP_2015}
introduced another short-range potential, termed as Voronoi 
soft-core potential, which they successfully used to analyze the
$^2\Pi_u$ resonance of CO$_2^-$.

Present analysis of the Gaussian perturbation potential
(\ref{eq-Gauss1}) will be carried out for expectedly simpler
problems - atomic shape resonances of beryllium and magnesium.
Both atoms are known to possess a $p$-wave shape resonance
very close to the elastic threshold.
While in the case of the Mg$^-$ the agreement between the 
available computed resonance parameters
\cite{Hunt_Moiseiwitsch_JPB_1970,Krylstedt_REB_JPB_1987,Kim_Greene_JPB_1989}
and the experimental data
\cite{Burrow_MC_JPB_1976,Buckman_Clark_RMP_1994} is quite good,
the situation is very different for the beryllium atom.
There has been a great number of theoretical studies
\cite{Kurtz_Ohrn_1979,Kurtz_Jordan_JPB_1981,McCurdy_RDL_JCP_1980,Rescigno_MO_PRA_1978,McNutt_McCurdy_PRA_1983,Krylstedt_EB_JPB_1988,Zhou_Ernzerhof_JPCL_2012,Venka_MJ_TCA_2000,Samanta_Yeager_JPCB_2008,Samanta_Yeager_IJQC_2010,Tsednee_LY_PRA_2015,Zatsarinny_BFB_JPB_2016}
aiming to numerically characterize Be$^-$ 2$s^2\varepsilon p\; ^2P$
resonance, with various levels of success. Table III in Ref.
\cite{Tsednee_LY_PRA_2015} clearly summarizes that the theory of the last
four decades predicts the resonance position between 0.1 and 1.2 eV
and the resonance width between 0.1 and 1.7 eV. Even the most recent
calculations differ by about a factor of 3 for the two resonant parameters.
Moreover, there are no experimental data available for the Be$^-$ resonance
that could narrow the spread of all the available theoretical predictions.

Convergence patters shown in Refs. 
\cite{Tsednee_LY_PRA_2015,Zatsarinny_BFB_JPB_2016} demonstrate that the
Be$^-$ resonance may be very sensitive to an accurate description of the
electronic correlation energy. Therefore, in the present study we employ
coupled-clusters (CCSD-T) and full configuration interaction (FCI) methods
for the perturbed Be$^-$ electron affinities that will be then continued
the complex plane by the RAC method. The basic ideas of the RAC method 
are given in the Sec.~\ref{sec-rac}. Quick summary of the quantum chemistry
details is presented in Sec.~\ref{sec-qchem}.
In Sec.~\ref{sec-res} we analyze the stability
and accuracy of the RAC method with the Gaussian perturbation potential
(\ref{eq-Gauss1}). Then conclusions follow.

\section{\label{sec-rac}RAC method}

The RAC method represents a very simple method for calculation of resonance 
energies and widths which embraces all known analytical features of coupling 
constant $\lambda(\kappa)$ near the zero energy \cite{JIR_JCP_2015}.
The method works as follows:

\begin{itemize}
\item The atom or molecule is perturbed by an attractive interaction $V$
multiplied by a real constant $\lambda$
\beq
\rm H_{neutral} \rightarrow H_{neutral} + \lambda V\;,
\eeq
and bound states energies $E^N_i$ of the neutral state are calculated for 
a set of values $\lambda_i$.

\item The same procedure is carried out for the corresponding negative ion
\beq
\rm H_{ion} \rightarrow H_{ion} + \lambda V\;,
\eeq
where the bound state energies $E^I_i$ are calculated for the same values of $\lambda_i$.
\item Both energies are subtracted forming the electron affinity
in the presence of the perturbation potential $V$
\beq
E^N_i - E^I_i = E_i = \kappa_i^2 \;.
\eeq
\end{itemize}
The new set of data points $\{\kappa_i,\lambda_i\}$ is then used to fit the function 
\beq
\label{eq-PA31}
\lambda(\kappa)=\lambda_0\frac{(\kappa^2+2\alpha^2\kappa+\alpha^4+\beta^2)
(1+\delta^2\kappa)}{\alpha^4+\beta^2+
\kappa(2\alpha^2+\delta^2(\alpha^4+\beta^2))}.
\eeq
It is represented as a Pad\'{e} 3/1 function and it defines the level of complexity of
the pole behavior at the low bound or continuum energies. We term it as 
RAC [3/1] method. The origin of its form and the fit formulae for 
[2/1], [3/2], and [4/2] methods can
be found in Ref.~\cite{JIR_EPJD_2016}.
The parameters of the [3/1] fit, namely $\alpha, \beta, \delta$ and $\lambda_0$ 
are found by minimizing the $\chi^2$ functional
\beq
\label{eq-chisqr}
\chi^2=\frac{1}{N}\sum_{i=1}^N\frac{1}{\varepsilon_i^2}\left|
       \lambda(\kappa_i)-\lambda_i\right|^2,
\eeq
where $N$ denotes the number of the points used, while $\kappa_i$ and $\lambda_i$ 
are the input data.
Once an accurate fit is found, only the parameters $\alpha$ and $\beta$
determine the resonance energy
\beq
E_r=\beta^2-\alpha^4\;,
\eeq
and the resonance width
\beq
\Gamma=4\beta\alpha^2\;.
\eeq

Role of the parameter $\delta$ is to describe a virtual state with $E_v=-1/\delta^4$. 
Even in the case the studied system does not possess a virtual state this parameter
represents a cumulative effect of the other resonances and 
other poles not explicitly included in the model.
The weights $\varepsilon_i$ (accuracy of the data) in Eq.~(\ref{eq-chisqr}) are generally unknown. 
The calculation can be routinely performed with constant $\varepsilon_i$ = 1 
or, if an importance of the data points closest to the origin needs to be stressed,
increasing weights sequence (e.g. $\varepsilon_i$ = i) can be used.

The RAC method has been recently critically evaluated by
\citeauthor{White_HGMC_JCP_2017} \cite{White_HGMC_JCP_2017}. Authors tested
three types of the perturbation potential
\begin{eqnarray}
V(r) &=& -\frac{\lambda}{r},\\
\label{eq-cg-pot}
V(r) &=& -\lambda \frac{e^{-\alpha r^2}}{r},\\
\label{eq-g-pot}
V(r) &=& -\lambda e^{-\alpha r^2}\;,
\end{eqnarray}
and they suggested that the attenuated Coulomb potential (\ref{eq-cg-pot}) is
the best choice out of the three options and the Gaussian potential
(\ref{eq-g-pot}) does not represent a good choice for the RAC method. 
All these potentials are
easily implemented into the standard quantum chemistry codes. The aim of the
present contribution is twofold:
\begin{itemize}
\item to explore application of the Gaussian-type perturbation 
and to find its parameters that allow accurate extraction of the resonance
data with the RAC method
\item to demonstrate that the RAC method can be applied with success to
low-lying atomic shape resonances
\end{itemize} 
Before applying the RAC method one must consider two important issues.
\begin{itemize}
\item [1.] First is a choice of the perturbation potential, i.e. in the present
context the choice of the exponent $\alpha$ in Eq.~(\ref{eq-g-pot}).
Presently there exist no general rule, no guide that helps us to choose
the perturbation potential. Therefore, it is necessary to perform
calculations for a set of values of the parameter $\alpha$ to find
an optimal choice. If the optimal range of values is found, it is
reasonable to expect that the obtained resonance data should stabilize
in such a range, because the exact function $\lambda(\kappa)$ gives
the same resonant data for every choice of the perturbation potential.
Since the present [3/1] RAC function is only approximative, one
can only expect an existence of a plateau that gives approximative values
of the resonance parameters.
\item [2.] The RAC method represents essentially a low-energy approximation
to the exact function $\lambda(\kappa)$. It is therefore obvious that
the method should be used in a range of energies (or momenta)
limited by some maximal energy $E_M$. Our empirical experience shows that
$E_m \sim 8 E_r$ ($E_r$ is the sought resonance energy) gives a reasonable
estimate for the range of energies.
\end{itemize}

\section{\label{sec-qchem}Electron affinities}

{\it Ab initio} calculations for the electron affinities $E_i(\lambda_i)$
in presence of the external Gaussian field (\ref{eq-g-pot}) were carried
out using the CCSD-T 
\cite{Knowles_HW_CC_JCP_1993,Deegan_Knowles_CPL_1994} and FCI methods
as implemented in MOLPRO 10 package of
quantum-chemistry programs \cite{MOLPRO10}. Core of the basis set employs
Dunning's augmented correlation-consistent basis of quadruple-zeta quality
aug-cc-pVQZ \cite{Prascher_WOKDW_TCA_2011} for both atoms, Be and Mg. This basis set
was additionally extended, in an even-tempered fashion, by 2
($s$, $d$, $f$, $g$)-type functions and 6 $p$-type functions.
\begin{figure}[thb]
\includegraphics[width=0.8\textwidth]{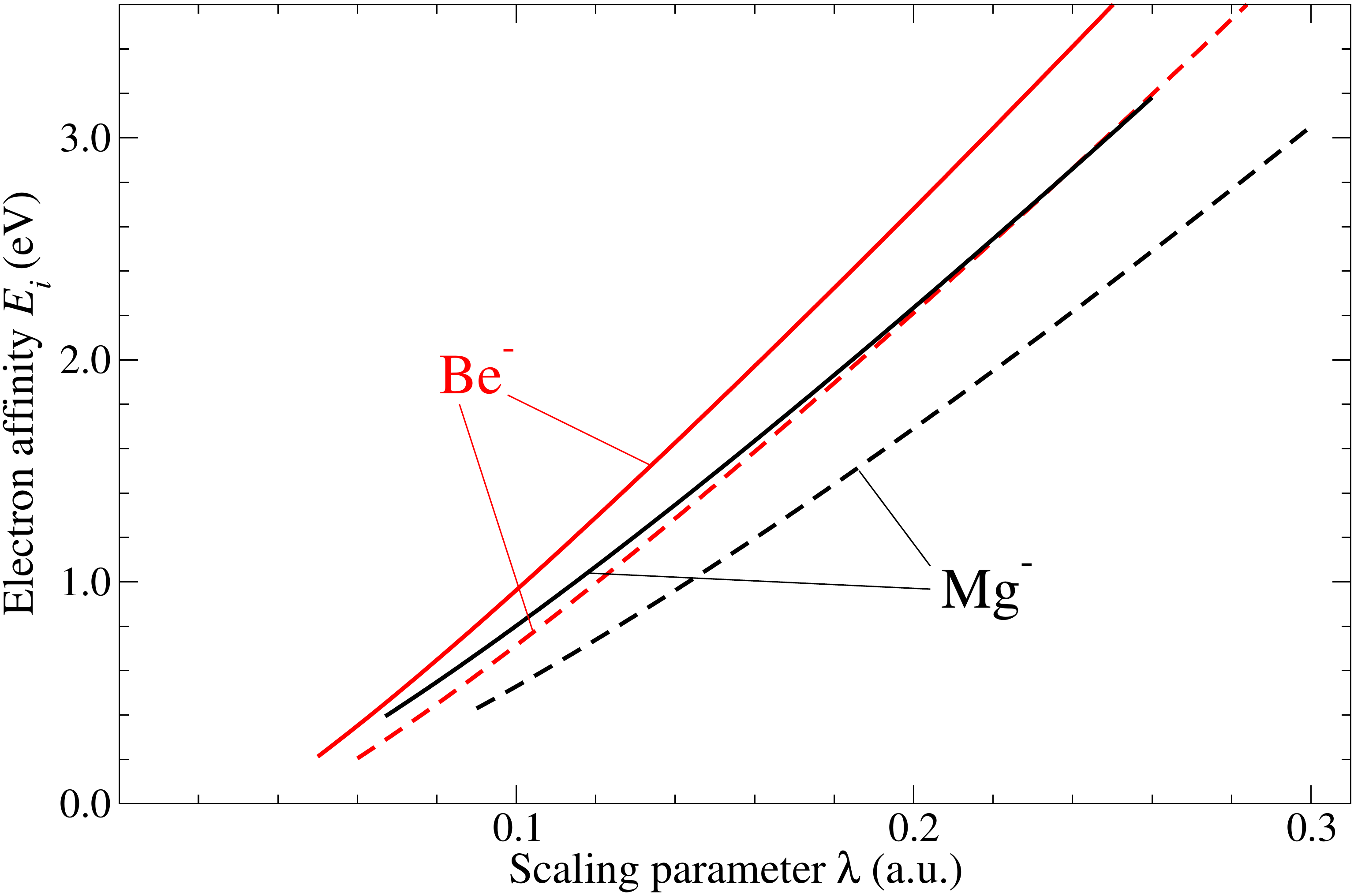}
\caption{\label{fig-eas}
(Color online) Electron affinities of Be$^-$ and Mg$^-$ ions under the
influence of the perturbation potential (\ref{eq-g-pot}). Full lines
are shown for the exponent $\alpha$ = 0.025, while the broken lines
are for $\alpha$ = 0.035. Red color (light gray) describes the Be$^-$
ion and the black color is for the Mg$^-$.
}
\end{figure}
Calculations
for the neutral atoms and corresponding negative ions used the same
basis sets and the same correlation methods (CCSD-T or FCI). 
Typical dependence of the electron affinities on 
external field (\ref{eq-g-pot}) is shown in Fig.~\ref{fig-eas}
for both negative ions, Be$^-$ and Mg$^-$, and in the range of energies
used for the present analytic continuation. Fig.~\ref{fig-eas} yields
the following observations:
\begin{itemize}
\item As expected, the weaker perturbation potential with 
      $\alpha$ = 0.035 requires 
      a stronger scaling parameter $\lambda$ to achieve the same
      binding negative ion energies as the perturbation with
      $\alpha$ = 0.025.
\item Surprisingly, a larger scaling parameter (stronger perturbation)
      is necessary to bind the Mg$^-$ resonance that lies closer
      to the zero when compared to the Be$^-$ resonance (as will be seen
      below). Such behavior may be caused by the spatial extent of the Mg$^-$
      3$p$ resonant wave function when compared to the reach of the
      2$p$ wave function of the Be$^-$ ion.
\item The lowest binding energies are not included in the continuation
      input data because of the difficulties we encountered while using
      the quantum chemistry software. Hartree-Fock method
      is known to destabilize in very diffused basis sets, however
      low binding energies are inaccurate if a more compact
      basis is used.
\end{itemize}

Most of the present results were obtained with the CCSD-T method. However, once
the the optimal exponent $\alpha$ (see the Sec.~\ref{sec-res}) was found
for the beryllium atom, the affinity curve shown in Fig.~\ref{fig-eas} was
also recomputed with the expensive FCI method and the basis as described above.

\section{\label{sec-res}Results}

\begin{figure}[thb]
\includegraphics[width=0.8\textwidth]{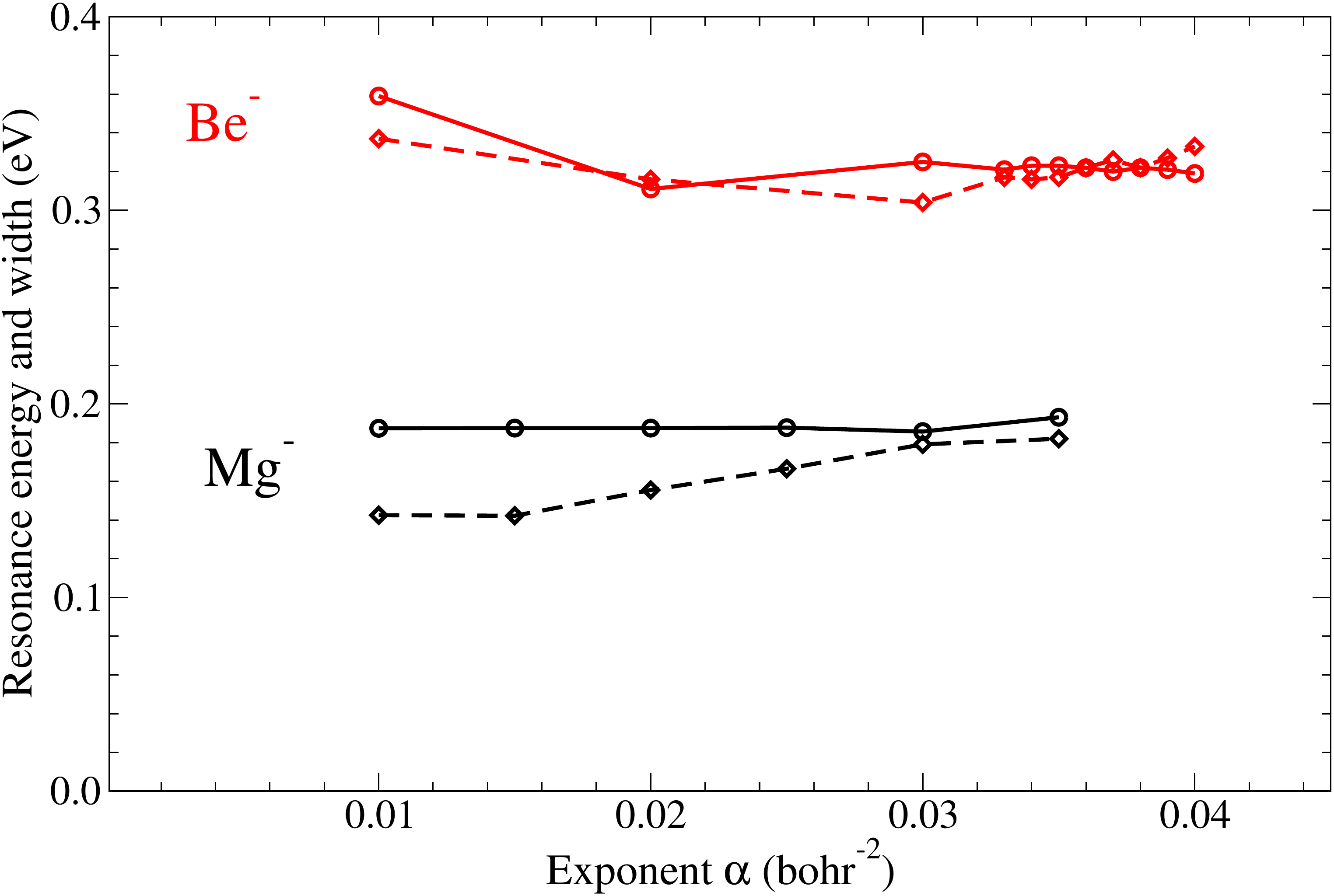}
\caption{\label{fig-alpha}
(Color online) Resonance energy $E_r$ (circles connected full lines) and the
resonance width $\Gamma$ (diamonds connected with dashed lines) as functions
of $\alpha$ parameter of the perturbation potential.
}
\end{figure}

As discussed in Sec.~\ref{sec-rac}, our goal is to search for regions of stable
results with respect to the two optimization parameters. First is the range of the
input electron affinities defined by maximal affinity $E_M$. The second parameter,
the exponent $\alpha$ in Eq.~(\ref{eq-g-pot}) defines the shape of the perturbation
potential.
\begin{figure}[thb]
\includegraphics[width=0.8\textwidth]{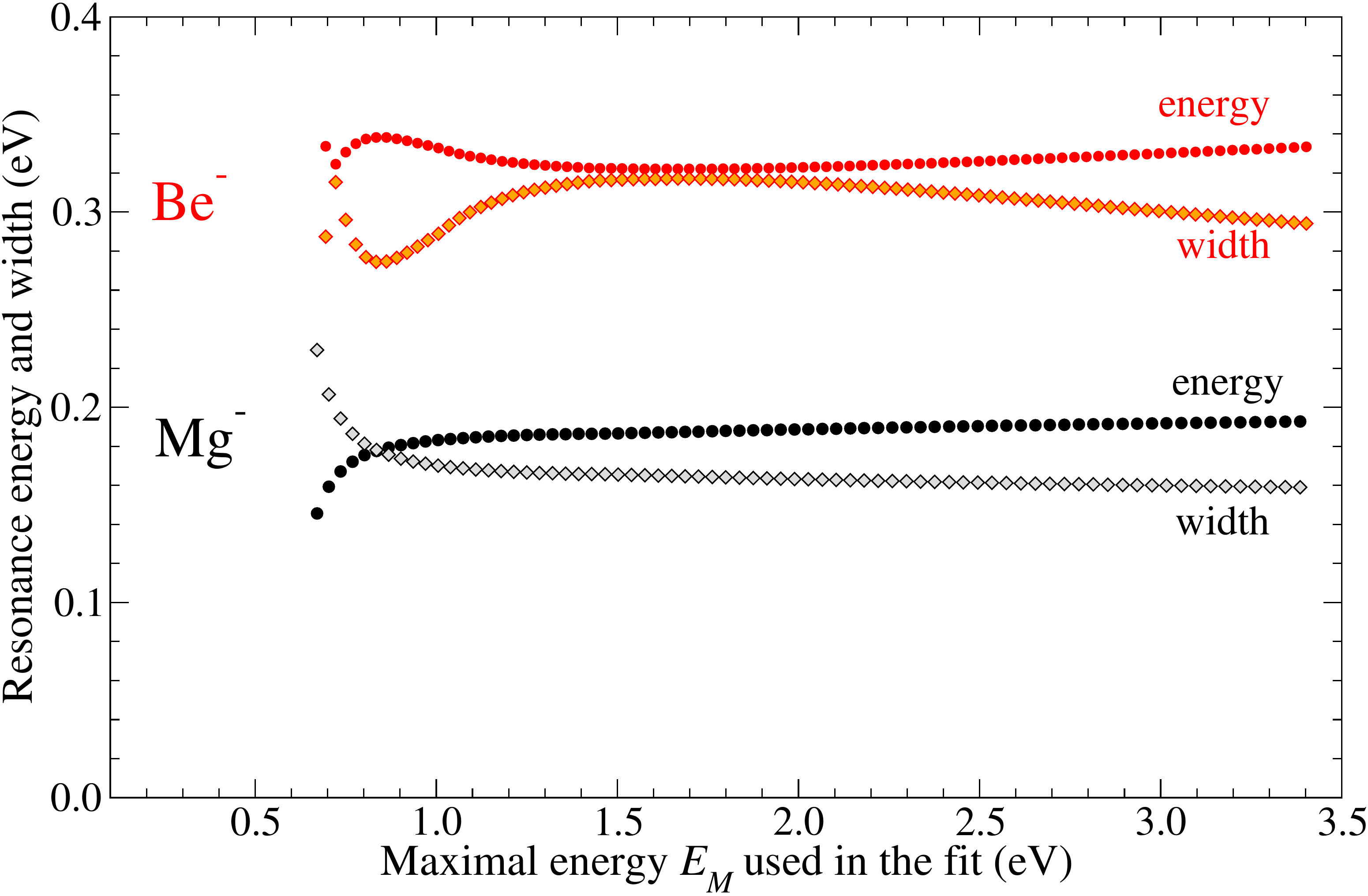}
\caption{\label{fig-emax}
(Color online) Resonance energy (shown as circles) and width (displayed as diamonds) 
calculated for Be$^-$ and Mg$^-$ as functions of the energy extent defined by 
the maximal energy $E_M$. The exponents $\alpha$ are fixed at $\alpha$ = 0.035 
for Be$^-$ and $\alpha$ = 0.025 for Mg$^-$.
}
\end{figure}
\begin{figure}[thb]
\includegraphics[width=0.8\textwidth]{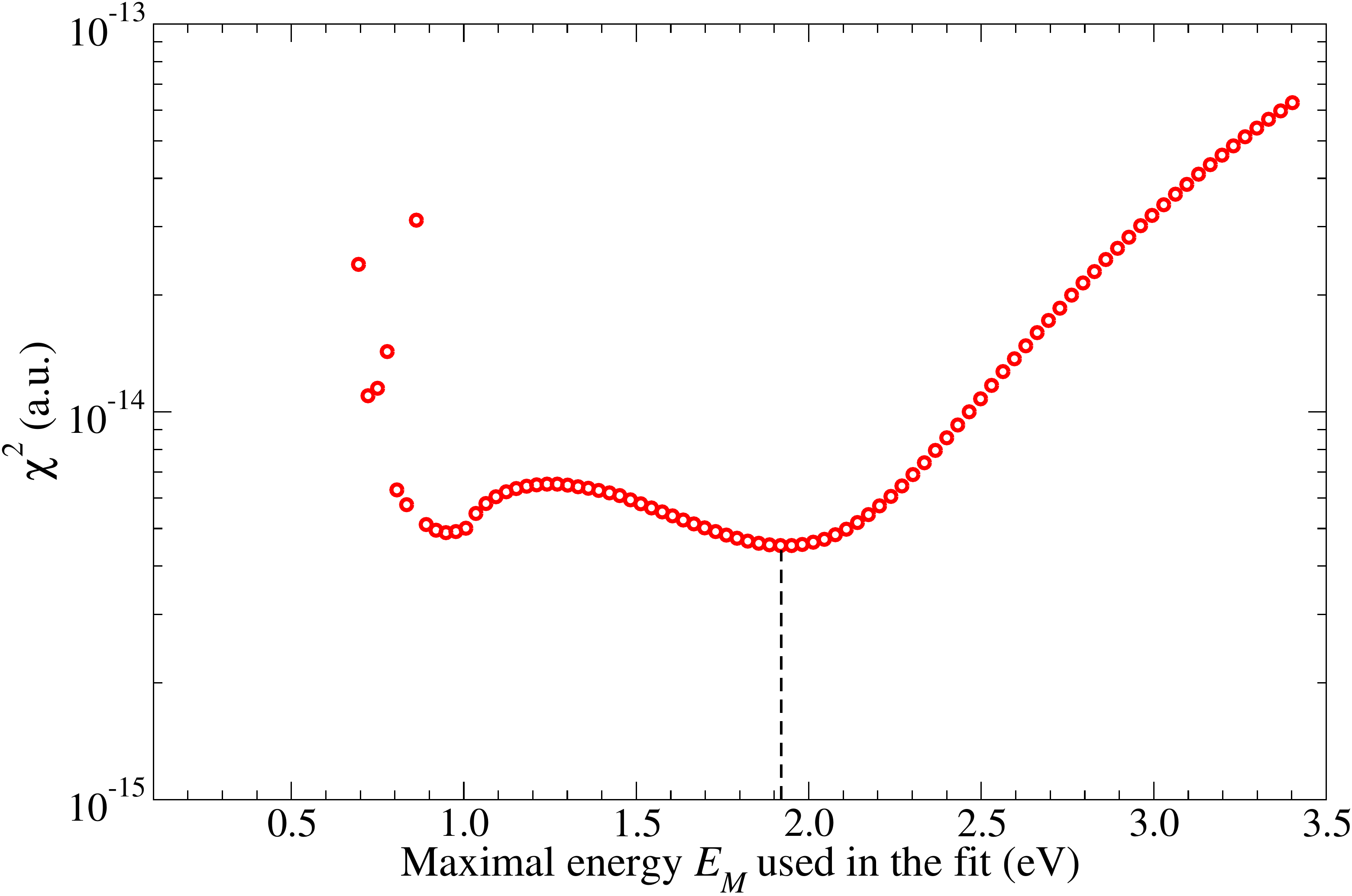}
\caption{\label{fig-chi2}
(Color online) Quality of the RAC fit for the resonance of Be$^-$
as a function of the maximal energy. Exponent $\alpha$ is fixed at 0.035
and the increasing weights set $\varepsilon$ = $i$ are used.
}
\end{figure}
Typical dependence of the resonance parameters on the maximal energy is shown
in Fig.~\ref{fig-emax} for the fixed $\alpha$ parameters.
It is clear that the stability is little worse for the Be$^-$ ion when compared
to Mg$^-$ ion. However, it is possible to narrow the spread of the obtained
resonance data by considering the value of $\chi^2$ defined by 
Eq.~(\ref{eq-chisqr}). 
Fig.~\ref{fig-chi2} shows the dependence of $\chi^2$ quantity on the
maximal energy $E_M$. A pronounced minimum at $E_M$ = 1.92 eV is clearly
visible. This allows an application of a condition of the best fit. Such a
restriction leads to a well defined $E_M$ for each choice of the perturbation
parameter $\alpha$ producing a data sets shown in Fig~\ref{fig-alpha}.
For beryllium the resonance position and width stabilizes for 
$\alpha >$~0.02. The best fit is obtained for $\alpha$ = 0.035 resulting in
$E_r$ = 0.323 eV and $\Gamma$ = 0.317 eV. In order to estimate accuracy
of the correlation energy provided by the CCSD-T method we also recomputed this
final results with the FCI method. The FCI affinities yield 
$E_r$ = 0.282 eV and $\Gamma$ = 0.316 eV.
Detailed summary of the available theoretical results 
for the Be$^-$ resonance was presented in
Tab.~III of Ref.~\cite{Tsednee_LY_PRA_2015}.
A comparison with the most recent computations will be given in 
Sec.~\ref{sec-conc}.

In
case of magnesium ion the resonance energy is very stable over the whole
range of examined perturbation parameters $\alpha$. However, the width exhibits
a weak dependence on the exponent $\alpha$. This feature may indicate that the 
low-order RAC method is inadequate for the Mg$^-$ resonance. Nonetheless, the
best fit is obtained for $\alpha$ = 0.025, giving $E_r$ = 0.188 eV and 
$\Gamma$ = 0.167 eV. The available data for the Mg$^-$ resonance are summarized
in Tab.~\ref{tab-Mg}.
\begin{table}
\caption{\label{tab-Mg}
         Comparison of the available data for the resonance energy 
         $E_r$ and the resonance width $\Gamma$ for the 
         3s$^2\varepsilon$p $^2$P state of atomic magnesium.
}
\begin{center}
\begin{tabular}{lcc}
\hline
Method & Resonance energy $E_r$(eV)  & Resonance width $\Gamma$(eV) \\
\hline\hline
Model potential \cite{Hunt_Moiseiwitsch_JPB_1970} &  0.37   &    0.10\\
Model potential \cite{Kim_Greene_JPB_1989}        &  0.161  &    0.160\\
Complex rotation \cite{Krylstedt_REB_JPB_1987}    &  0.08   &    0.17\\
Stabilization \cite{Chao_FY_JCP_1990}             &  0.14   &    0.08\\
Complex SCF \cite{McCurdy_LM_JCP_1981}            &  0.50   &    0.54\\
Finite elements \cite{Gallup_PRA_2011}            &  0.159  &    0.12\\
Experiment \cite{Burrow_MC_JPB_1976}              &  0.15$\pm$0.03 &   $\sim$0.14\\
Recommended value \cite{Buckman_Clark_RMP_1994}   &  0.15   &    0.16\\
Present RAC                                       &  0.19   &    0.16\\
\hline
\end{tabular}

\end{center}
\end{table}
Presently computed resonance energy is about 40 meV higher that the experimental
value of \citeauthor{Burrow_MC_JPB_1976} 
\cite{Burrow_Comer_JPB_1985,Burrow_MC_JPB_1976}. Such a discrepancy may have several
possible reasons:
\begin{itemize}
\item[1.] The experimental resolution is about 30--40 meV \cite{Burrow_MC_JPB_1976}.
\item[2.] Discrepancy between the correlation energies of the CCSD-T and FCI methods
          and the present basis set is about 41 meV for the electron affinity of
          the beryllium atom. Similar difference can also be expected for the magnesium.
          Moreover,
          weaker stability of $\Gamma$ with respect to the perturbation potential 
          (shown in Fig.~\ref{fig-alpha})
          indicates that higher order continuation may be necessary.
\item[3.] The experimental resonance energy \cite{Burrow_MC_JPB_1976} was determined
           from the maximum of the measured cross section, whereas present method
           defines the resonance energy from a pole of the $S$-matrix. The two definitions
           give similar results for a narrow resonance ($\Gamma < E_r$), but for
           for a broader resonance ($\Gamma \geq E_r$), as in the present case,
           the results may differ.
\end{itemize}

\section{\label{sec-conc}Conclusions}

Present study confirms the observations of \citeauthor{White_HGMC_JCP_2017} 
\cite{White_HGMC_JCP_2017} in which the authors state that the Gaussian
perturbation potential is more difficult to apply than potentials possessing
the Coulomb singularity. It has been shown in the case of a model potential
\cite{White_HGMC_JCP_2017} that the trajectory of the resonant pole is more
complicated for the Gaussian perturbation. In the present study we have shown
that in order to obtain stable results, the RAC method must be restricted 
to fairly low electron affinities and a careful analysis of the results with
respect to the width of the perturbation potential must be carried out.

Such procedure allowed us to apply the RAC method to one of the remaining
enigmas among shape resonances of small atoms, the 2$s^2\varepsilon p\; ^2P$
resonance of Be$^-$. To the best of our knowledge there are no experimental
data available for this resonance. Important role of the correlation energy
in this system creates a challenging task for the theory, albeit the fact
that Be$^-$ possess only 5 electrons. Consequently, about two dozens of
theoretical predictions (found in Ref.~\cite{Kurtz_Ohrn_1979,Kurtz_Jordan_JPB_1981,McCurdy_RDL_JCP_1980,Rescigno_MO_PRA_1978,McNutt_McCurdy_PRA_1983,Krylstedt_EB_JPB_1988,Zhou_Ernzerhof_JPCL_2012,Venka_MJ_TCA_2000,Samanta_Yeager_JPCB_2008,Samanta_Yeager_IJQC_2010,Tsednee_LY_PRA_2015,Zatsarinny_BFB_JPB_2016})
do not result in any kind of a consensus. 
Two methods with high level of correlation description, the CCSD-T and FCI methods, 
were applied
in the present study. While the position of the resonance shifts to the lower energies
by about 41 meV for the more accurate FCI method, the resonance width was found insensitive
to the correlation treatment.
Presently calculated FCI
resonant energy $E_r$ = 0.282 eV and width $\Gamma$ = 0.316 eV are in 
a good agreement with the complex CI results of
\citeauthor{McNutt_McCurdy_PRA_1983} \cite{McNutt_McCurdy_PRA_1983} that
predict the $E_r$ = 0.323 eV and $\Gamma$ = 0.296 eV. Recent scattering
calculations 
\cite{Zatsarinny_BFB_JPB_2016} determined the resonance with $E_r$ = 
$0.31 \pm 0.04$ eV and $\Gamma$ = $0.40 \pm 0.06$ eV again in a good 
agreement with the present results.
However, another set of recent calculations by
\citeauthor{Tsednee_LY_PRA_2015} \cite{Tsednee_LY_PRA_2015} place the
resonance at $E_r$ = 0.756 eV and $\Gamma$ = 0.874 eV.

In case of the 2$s^2\varepsilon p\; ^2P$ resonance of Mg$^-$ a comparison
with the experiment is available. Although, the present calculations determine
the resonance about 40~meV higher than the experiment 
\cite{Burrow_MC_JPB_1976}, they still exhibit the best agreement with the
experimental data among the {ab-initio} methods.

\begin{acknowledgments}
The contributions of R\v{C} were supported by
the Grant Agency of the Czech Republic (Grant No. GACR 18-02098S).
JH conducted this work with support of the Grant Agency of Czech Republic
(Grant No. GACR 16-17230S).
IP acknowledges support from the Grant Agency of the Czech Republic
(Grant No. GACR 17-14200S).
\end{acknowledgments}

\bibliographystyle{apsrev}
\bibliography{RES}

\begin{thebibliography}{50}
\expandafter\ifx\csname natexlab\endcsname\relax\def\natexlab#1{#1}\fi
\expandafter\ifx\csname bibnamefont\endcsname\relax
  \def\bibnamefont#1{#1}\fi
\expandafter\ifx\csname bibfnamefont\endcsname\relax
  \def\bibfnamefont#1{#1}\fi
\expandafter\ifx\csname citenamefont\endcsname\relax
  \def\citenamefont#1{#1}\fi
\expandafter\ifx\csname url\endcsname\relax
  \def\url#1{\texttt{#1}}\fi
\expandafter\ifx\csname urlprefix\endcsname\relax\def\urlprefix{URL }\fi
\providecommand{\bibinfo}[2]{#2}
\providecommand{\eprint}[2][]{\url{#2}}

\bibitem[{\citenamefont{Carelli et~al.}(2013)\citenamefont{Carelli, Satta,
  Grassi, and Gianturco}}]{Carelli_FAG_2013}
\bibinfo{author}{\bibfnamefont{F.}~\bibnamefont{Carelli}},
  \bibinfo{author}{\bibfnamefont{M.}~\bibnamefont{Satta}},
  \bibinfo{author}{\bibfnamefont{T.}~\bibnamefont{Grassi}}, \bibnamefont{and}
  \bibinfo{author}{\bibfnamefont{F.~A.} \bibnamefont{Gianturco}},
  \bibinfo{journal}{Astrophys. J.} \textbf{\bibinfo{volume}{774}},
  \bibinfo{pages}{97} (\bibinfo{year}{2013}).

\bibitem[{\citenamefont{van Dorp}(2012)}]{Dorp_PCCP_2012}
\bibinfo{author}{\bibfnamefont{W.~F.} \bibnamefont{van Dorp}},
  \bibinfo{journal}{Phys. Chem. Chem. Phys.} \textbf{\bibinfo{volume}{14}},
  \bibinfo{pages}{16753} (\bibinfo{year}{2012}).

\bibitem[{\citenamefont{Thorman et~al.}(2015)\citenamefont{Thorman, Kumar,
  Fairbrother, and Ing\'olfsson}}]{Thorman_Oddur_BJN_2015}
\bibinfo{author}{\bibfnamefont{R.~M.} \bibnamefont{Thorman}},
  \bibinfo{author}{\bibfnamefont{R.~T.~P.} \bibnamefont{Kumar}},
  \bibinfo{author}{\bibfnamefont{D.~H.} \bibnamefont{Fairbrother}},
  \bibnamefont{and}
  \bibinfo{author}{\bibfnamefont{O.}~\bibnamefont{Ing\'olfsson}},
  \bibinfo{journal}{Beilstein J. Nanotechnol.} \textbf{\bibinfo{volume}{6}},
  \bibinfo{pages}{1904} (\bibinfo{year}{2015}).

\bibitem[{\citenamefont{Pan et~al.}(2003)\citenamefont{Pan, Cloutier, Hunting,
  and Sanche}}]{Sanche_PRL_2003}
\bibinfo{author}{\bibfnamefont{X.}~\bibnamefont{Pan}},
  \bibinfo{author}{\bibfnamefont{P.}~\bibnamefont{Cloutier}},
  \bibinfo{author}{\bibfnamefont{D.}~\bibnamefont{Hunting}}, \bibnamefont{and}
  \bibinfo{author}{\bibfnamefont{L.}~\bibnamefont{Sanche}},
  \bibinfo{journal}{Phys. Rev. Lett.} \textbf{\bibinfo{volume}{90}},
  \bibinfo{pages}{208102} (\bibinfo{year}{2003}).

\bibitem[{\citenamefont{Zheng et~al.}(2006)\citenamefont{Zheng, Wagner, and
  Sanche}}]{Sanche_PRL_2006}
\bibinfo{author}{\bibfnamefont{Y.}~\bibnamefont{Zheng}},
  \bibinfo{author}{\bibfnamefont{J.~R.} \bibnamefont{Wagner}},
  \bibnamefont{and} \bibinfo{author}{\bibfnamefont{L.}~\bibnamefont{Sanche}},
  \bibinfo{journal}{Phys. Rev. Lett.} \textbf{\bibinfo{volume}{96}},
  \bibinfo{pages}{208101} (\bibinfo{year}{2006}).

\bibitem[{\citenamefont{Taylor and Hazi}(1976)}]{Taylor_Hazi_PRA_1976}
\bibinfo{author}{\bibfnamefont{H.~S.} \bibnamefont{Taylor}} \bibnamefont{and}
  \bibinfo{author}{\bibfnamefont{A.~U.} \bibnamefont{Hazi}},
  \bibinfo{journal}{Phys. Rev. A} \textbf{\bibinfo{volume}{14}},
  \bibinfo{pages}{2071} (\bibinfo{year}{1976}).

\bibitem[{\citenamefont{Hazi and Taylor}(1970)}]{Hazi_Taylor_PRA_1970}
\bibinfo{author}{\bibfnamefont{A.~U.} \bibnamefont{Hazi}} \bibnamefont{and}
  \bibinfo{author}{\bibfnamefont{H.~S.} \bibnamefont{Taylor}},
  \bibinfo{journal}{Phys. Rev. A} \textbf{\bibinfo{volume}{1}},
  \bibinfo{pages}{1109} (\bibinfo{year}{1970}).

\bibitem[{\citenamefont{Hazi et~al.}(1981)\citenamefont{Hazi, Rescigno, and
  Kurilla}}]{Hazi_Kurilla_PRA_1981}
\bibinfo{author}{\bibfnamefont{A.~U.} \bibnamefont{Hazi}},
  \bibinfo{author}{\bibfnamefont{T.~N.} \bibnamefont{Rescigno}},
  \bibnamefont{and} \bibinfo{author}{\bibfnamefont{M.}~\bibnamefont{Kurilla}},
  \bibinfo{journal}{Phys. Rev. A} \textbf{\bibinfo{volume}{23}},
  \bibinfo{pages}{1089} (\bibinfo{year}{1981}).

\bibitem[{\citenamefont{Frey and Simons}(1986)}]{Frey_Simons_JCP_1986}
\bibinfo{author}{\bibfnamefont{R.~F.} \bibnamefont{Frey}} \bibnamefont{and}
  \bibinfo{author}{\bibfnamefont{J.}~\bibnamefont{Simons}},
  \bibinfo{journal}{J. Chem. Phys.} \textbf{\bibinfo{volume}{84}},
  \bibinfo{pages}{4462} (\bibinfo{year}{1986}).

\bibitem[{\citenamefont{Moiseyev}(1998)}]{Moiseyev_PR_1998}
\bibinfo{author}{\bibfnamefont{N.}~\bibnamefont{Moiseyev}},
  \bibinfo{journal}{Phys. Rep.} \textbf{\bibinfo{volume}{302}},
  \bibinfo{pages}{212} (\bibinfo{year}{1998}).

\bibitem[{\citenamefont{McCurdy and
  Rescigno}(1978)}]{McCurdy_Rescigno_PRL_1978}
\bibinfo{author}{\bibfnamefont{C.~W.} \bibnamefont{McCurdy}} \bibnamefont{and}
  \bibinfo{author}{\bibfnamefont{T.~N.} \bibnamefont{Rescigno}},
  \bibinfo{journal}{Phys. Rev. Lett.} \textbf{\bibinfo{volume}{41}},
  \bibinfo{pages}{1364} (\bibinfo{year}{1978}).

\bibitem[{\citenamefont{Reinhardt}(1982)}]{Reinhardt_ARPC_1982}
\bibinfo{author}{\bibfnamefont{W.~P.} \bibnamefont{Reinhardt}},
  \bibinfo{journal}{Ann. Rev. Phys. Chem.} \textbf{\bibinfo{volume}{33}},
  \bibinfo{pages}{223} (\bibinfo{year}{1982}).

\bibitem[{\citenamefont{Riss and Meyer}(1993)}]{Riss_Meyer_JPB_1993}
\bibinfo{author}{\bibfnamefont{U.~V.} \bibnamefont{Riss}} \bibnamefont{and}
  \bibinfo{author}{\bibfnamefont{H.~D.} \bibnamefont{Meyer}},
  \bibinfo{journal}{J. Phys. B: Atom. Molec. Phys.}
  \textbf{\bibinfo{volume}{26}}, \bibinfo{pages}{4503} (\bibinfo{year}{1993}).

\bibitem[{\citenamefont{Feuerbacher et~al.}(2003)\citenamefont{Feuerbacher,
  Sommerfeld, Santra, and Cederbaum}}]{Feurbacher_Cederbaum_JCP_2003}
\bibinfo{author}{\bibfnamefont{S.}~\bibnamefont{Feuerbacher}},
  \bibinfo{author}{\bibfnamefont{T.}~\bibnamefont{Sommerfeld}},
  \bibinfo{author}{\bibfnamefont{R.}~\bibnamefont{Santra}}, \bibnamefont{and}
  \bibinfo{author}{\bibfnamefont{L.~S.} \bibnamefont{Cederbaum}},
  \bibinfo{journal}{The Journal of Chemical Physics}
  \textbf{\bibinfo{volume}{118}}, \bibinfo{pages}{6188} (\bibinfo{year}{2003}).

\bibitem[{\citenamefont{Kukulin and
  Krasnopolsky}(1977)}]{Kukulin_Krasnopolsky_JPA_1977}
\bibinfo{author}{\bibfnamefont{V.~I.} \bibnamefont{Kukulin}} \bibnamefont{and}
  \bibinfo{author}{\bibfnamefont{V.~M.} \bibnamefont{Krasnopolsky}},
  \bibinfo{journal}{J. Phys. A: Math. Gen.} \textbf{\bibinfo{volume}{10}},
  \bibinfo{pages}{L33} (\bibinfo{year}{1977}).

\bibitem[{\citenamefont{Krasnopolsky and
  Kukulin}(1978)}]{Krasnopolsky_Kukulin_PL_1978}
\bibinfo{author}{\bibfnamefont{V.~M.} \bibnamefont{Krasnopolsky}}
  \bibnamefont{and} \bibinfo{author}{\bibfnamefont{V.~I.}
  \bibnamefont{Kukulin}}, \bibinfo{journal}{Phys. Lett. A}
  \textbf{\bibinfo{volume}{69}}, \bibinfo{pages}{251} (\bibinfo{year}{1978}).

\bibitem[{\citenamefont{Kukulin et~al.}(1988)\citenamefont{Kukulin,
  Krasnopolsky, and Hor\'{a}\v{c}ek}}]{KKH_book}
\bibinfo{author}{\bibfnamefont{V.~I.} \bibnamefont{Kukulin}},
  \bibinfo{author}{\bibfnamefont{V.~M.} \bibnamefont{Krasnopolsky}},
  \bibnamefont{and}
  \bibinfo{author}{\bibfnamefont{J.}~\bibnamefont{Hor\'{a}\v{c}ek}},
  \emph{\bibinfo{title}{Theory of Resonances: Principles and Applications}}
  (\bibinfo{publisher}{Kluwer Academic Publishers},
  \bibinfo{address}{Dordrecht/Boston/London}, \bibinfo{year}{1988}).

\bibitem[{\citenamefont{Hor\'a\v{c}ek et~al.}(2010)\citenamefont{Hor\'a\v{c}ek,
  Mach, and Urban}}]{Horacek_MU_PRA_2010}
\bibinfo{author}{\bibfnamefont{J.}~\bibnamefont{Hor\'a\v{c}ek}},
  \bibinfo{author}{\bibfnamefont{P.}~\bibnamefont{Mach}}, \bibnamefont{and}
  \bibinfo{author}{\bibfnamefont{J.}~\bibnamefont{Urban}},
  \bibinfo{journal}{Phys. Rev. A} \textbf{\bibinfo{volume}{82}},
  \bibinfo{pages}{032713} (\bibinfo{year}{2010}).

\bibitem[{\citenamefont{White et~al.}(2017)\citenamefont{White, Head-Gordon,
  and McCurdy}}]{White_HGMC_JCP_2017}
\bibinfo{author}{\bibfnamefont{A.~F.} \bibnamefont{White}},
  \bibinfo{author}{\bibfnamefont{M.}~\bibnamefont{Head-Gordon}},
  \bibnamefont{and} \bibinfo{author}{\bibfnamefont{C.~W.}
  \bibnamefont{McCurdy}}, \bibinfo{journal}{J. Chem. Phys.}
  \textbf{\bibinfo{volume}{146}}, \bibinfo{pages}{044112}
  (\bibinfo{year}{2017}).

\bibitem[{\citenamefont{Hor\'{a}\v{c}ek
  et~al.}(2014)\citenamefont{Hor\'{a}\v{c}ek, Paidarov\'{a}, and
  \v{C}ur\'{\i}k}}]{JIR_JPCA_2014}
\bibinfo{author}{\bibfnamefont{J.}~\bibnamefont{Hor\'{a}\v{c}ek}},
  \bibinfo{author}{\bibfnamefont{I.}~\bibnamefont{Paidarov\'{a}}},
  \bibnamefont{and}
  \bibinfo{author}{\bibfnamefont{R.}~\bibnamefont{\v{C}ur\'{\i}k}},
  \bibinfo{journal}{J. Phys. Chem. A} \textbf{\bibinfo{volume}{118}},
  \bibinfo{pages}{6536} (\bibinfo{year}{2014}).

\bibitem[{\citenamefont{Sommerfeld et~al.}(2017)\citenamefont{Sommerfeld,
  Melugin, Hamal, and Ehara}}]{Sommerfeld_MHPEM_JCTC_2017}
\bibinfo{author}{\bibfnamefont{T.}~\bibnamefont{Sommerfeld}},
  \bibinfo{author}{\bibfnamefont{J.~B.} \bibnamefont{Melugin}},
  \bibinfo{author}{\bibfnamefont{P.}~\bibnamefont{Hamal}}, \bibnamefont{and}
  \bibinfo{author}{\bibfnamefont{M.}~\bibnamefont{Ehara}}, \bibinfo{journal}{J.
  Chem. Theory Comput.} \textbf{\bibinfo{volume}{13}}, \bibinfo{pages}{2550}
  (\bibinfo{year}{2017}).

\bibitem[{\citenamefont{Papp et~al.}(2013)\citenamefont{Papp, \v{S}.
  Matej\v{c}\'{\i}k, Mach, Urban, Paidarov\'{a}, and
  Hor\'a\v{c}ek}}]{Papp_Horacek_CP_2013}
\bibinfo{author}{\bibfnamefont{P.}~\bibnamefont{Papp}},
  \bibinfo{author}{\bibnamefont{\v{S}. Matej\v{c}\'{\i}k}},
  \bibinfo{author}{\bibfnamefont{P.}~\bibnamefont{Mach}},
  \bibinfo{author}{\bibfnamefont{J.}~\bibnamefont{Urban}},
  \bibinfo{author}{\bibfnamefont{I.}~\bibnamefont{Paidarov\'{a}}},
  \bibnamefont{and}
  \bibinfo{author}{\bibfnamefont{J.}~\bibnamefont{Hor\'a\v{c}ek}},
  \bibinfo{journal}{Chem. Phys.} \textbf{\bibinfo{volume}{418}},
  \bibinfo{pages}{8} (\bibinfo{year}{2013}).

\bibitem[{\citenamefont{\v{C}ur\'{\i}k
  et~al.}(2016)\citenamefont{\v{C}ur\'{\i}k, Paidarov\'{a}, and
  Hor\'{a}\v{c}ek}}]{JIR_EPJD_2016}
\bibinfo{author}{\bibfnamefont{R.}~\bibnamefont{\v{C}ur\'{\i}k}},
  \bibinfo{author}{\bibfnamefont{I.}~\bibnamefont{Paidarov\'{a}}},
  \bibnamefont{and}
  \bibinfo{author}{\bibfnamefont{J.}~\bibnamefont{Hor\'{a}\v{c}ek}},
  \bibinfo{journal}{Europ. Phys. J. D} \textbf{\bibinfo{volume}{70}},
  \bibinfo{pages}{146} (\bibinfo{year}{2016}).

\bibitem[{\citenamefont{Hor\'{a}\v{c}ek
  et~al.}(2015)\citenamefont{Hor\'{a}\v{c}ek, Paidarov\'{a}, and
  \v{C}ur\'{\i}k}}]{JIR_JCP_2015}
\bibinfo{author}{\bibfnamefont{J.}~\bibnamefont{Hor\'{a}\v{c}ek}},
  \bibinfo{author}{\bibfnamefont{I.}~\bibnamefont{Paidarov\'{a}}},
  \bibnamefont{and}
  \bibinfo{author}{\bibfnamefont{R.}~\bibnamefont{\v{C}ur\'{\i}k}},
  \bibinfo{journal}{J. Chem. Phys.} \textbf{\bibinfo{volume}{143}},
  \bibinfo{pages}{184102} (\bibinfo{year}{2015}).

\bibitem[{\citenamefont{Sommerfeld and
  Ehara}(2015)}]{Sommerfeld_Ehara_JCP_2015}
\bibinfo{author}{\bibfnamefont{T.}~\bibnamefont{Sommerfeld}} \bibnamefont{and}
  \bibinfo{author}{\bibfnamefont{M.}~\bibnamefont{Ehara}}, \bibinfo{journal}{J.
  Chem. Phys.} \textbf{\bibinfo{volume}{142}}, \bibinfo{pages}{034105}
  (\bibinfo{year}{2015}).

\bibitem[{\citenamefont{Hunt and
  Moiseiwitsch}(1970)}]{Hunt_Moiseiwitsch_JPB_1970}
\bibinfo{author}{\bibfnamefont{J.}~\bibnamefont{Hunt}} \bibnamefont{and}
  \bibinfo{author}{\bibfnamefont{B.~L.} \bibnamefont{Moiseiwitsch}},
  \bibinfo{journal}{J. Phys. B: Atom. Molec. Phys.}
  \textbf{\bibinfo{volume}{3}}, \bibinfo{pages}{892} (\bibinfo{year}{1970}).

\bibitem[{\citenamefont{Krylstedt et~al.}(1987)\citenamefont{Krylstedt, Rittby,
  Elander, and Brandas}}]{Krylstedt_REB_JPB_1987}
\bibinfo{author}{\bibfnamefont{P.}~\bibnamefont{Krylstedt}},
  \bibinfo{author}{\bibfnamefont{M.}~\bibnamefont{Rittby}},
  \bibinfo{author}{\bibfnamefont{N.}~\bibnamefont{Elander}}, \bibnamefont{and}
  \bibinfo{author}{\bibfnamefont{E.}~\bibnamefont{Brandas}},
  \bibinfo{journal}{J. Phys. B: Atom. Molec. Phys.}
  \textbf{\bibinfo{volume}{20}}, \bibinfo{pages}{1295} (\bibinfo{year}{1987}).

\bibitem[{\citenamefont{Kim and Greene}(1989)}]{Kim_Greene_JPB_1989}
\bibinfo{author}{\bibfnamefont{L.}~\bibnamefont{Kim}} \bibnamefont{and}
  \bibinfo{author}{\bibfnamefont{C.~H.} \bibnamefont{Greene}},
  \bibinfo{journal}{J. Phys. B: Atom. Molec. Phys.}
  \textbf{\bibinfo{volume}{22}}, \bibinfo{pages}{L175} (\bibinfo{year}{1989}).

\bibitem[{\citenamefont{Burrow et~al.}(1976)\citenamefont{Burrow, Michejda, and
  Comer}}]{Burrow_MC_JPB_1976}
\bibinfo{author}{\bibfnamefont{P.~D.} \bibnamefont{Burrow}},
  \bibinfo{author}{\bibfnamefont{J.~A.} \bibnamefont{Michejda}},
  \bibnamefont{and} \bibinfo{author}{\bibfnamefont{J.}~\bibnamefont{Comer}},
  \bibinfo{journal}{J. Phys. B: Atom. Molec. Phys.}
  \textbf{\bibinfo{volume}{9}}, \bibinfo{pages}{3225} (\bibinfo{year}{1976}).

\bibitem[{\citenamefont{Buckman and Clark}(1994)}]{Buckman_Clark_RMP_1994}
\bibinfo{author}{\bibfnamefont{S.~J.} \bibnamefont{Buckman}} \bibnamefont{and}
  \bibinfo{author}{\bibfnamefont{C.~W.} \bibnamefont{Clark}},
  \bibinfo{journal}{Rev. Mod. Phys.} \textbf{\bibinfo{volume}{66}},
  \bibinfo{pages}{539} (\bibinfo{year}{1994}).

\bibitem[{\citenamefont{Kurtz and \"Ohrn}(1979)}]{Kurtz_Ohrn_1979}
\bibinfo{author}{\bibfnamefont{H.~A.} \bibnamefont{Kurtz}} \bibnamefont{and}
  \bibinfo{author}{\bibfnamefont{Y.}~\bibnamefont{\"Ohrn}},
  \bibinfo{journal}{Phys. Rev. A} \textbf{\bibinfo{volume}{19}},
  \bibinfo{pages}{43} (\bibinfo{year}{1979}).

\bibitem[{\citenamefont{Kurtz and Jordan}(1981)}]{Kurtz_Jordan_JPB_1981}
\bibinfo{author}{\bibfnamefont{H.~A.} \bibnamefont{Kurtz}} \bibnamefont{and}
  \bibinfo{author}{\bibfnamefont{K.~D.} \bibnamefont{Jordan}},
  \bibinfo{journal}{J. Phys. B: Atom. Molec. Phys.}
  \textbf{\bibinfo{volume}{14}}, \bibinfo{pages}{4361} (\bibinfo{year}{1981}).

\bibitem[{\citenamefont{McCurdy et~al.}(1980)\citenamefont{McCurdy, Rescigno,
  Davidson, and Lauderdale}}]{McCurdy_RDL_JCP_1980}
\bibinfo{author}{\bibfnamefont{C.~W.} \bibnamefont{McCurdy}},
  \bibinfo{author}{\bibfnamefont{T.~N.} \bibnamefont{Rescigno}},
  \bibinfo{author}{\bibfnamefont{E.~R.} \bibnamefont{Davidson}},
  \bibnamefont{and} \bibinfo{author}{\bibfnamefont{J.~G.}
  \bibnamefont{Lauderdale}}, \bibinfo{journal}{J. Chem. Phys.}
  \textbf{\bibinfo{volume}{73}}, \bibinfo{pages}{3268} (\bibinfo{year}{1980}).

\bibitem[{\citenamefont{Rescigno et~al.}(1978)\citenamefont{Rescigno, McCurdy,
  and Orel}}]{Rescigno_MO_PRA_1978}
\bibinfo{author}{\bibfnamefont{T.~N.} \bibnamefont{Rescigno}},
  \bibinfo{author}{\bibfnamefont{C.~W.} \bibnamefont{McCurdy}},
  \bibnamefont{and} \bibinfo{author}{\bibfnamefont{A.~E.} \bibnamefont{Orel}},
  \bibinfo{journal}{Phys. Rev. A} \textbf{\bibinfo{volume}{17}},
  \bibinfo{pages}{1931} (\bibinfo{year}{1978}).

\bibitem[{\citenamefont{McNutt and McCurdy}(1983)}]{McNutt_McCurdy_PRA_1983}
\bibinfo{author}{\bibfnamefont{J.~F.} \bibnamefont{McNutt}} \bibnamefont{and}
  \bibinfo{author}{\bibfnamefont{C.~W.} \bibnamefont{McCurdy}},
  \bibinfo{journal}{Phys. Rev. A} \textbf{\bibinfo{volume}{27}},
  \bibinfo{pages}{132} (\bibinfo{year}{1983}).

\bibitem[{\citenamefont{Krylstedt et~al.}(1988)\citenamefont{Krylstedt,
  Elander, and Brandas}}]{Krylstedt_EB_JPB_1988}
\bibinfo{author}{\bibfnamefont{P.}~\bibnamefont{Krylstedt}},
  \bibinfo{author}{\bibfnamefont{N.}~\bibnamefont{Elander}}, \bibnamefont{and}
  \bibinfo{author}{\bibfnamefont{E.}~\bibnamefont{Brandas}},
  \bibinfo{journal}{J. Phys. B: Atom. Molec. Phys.}
  \textbf{\bibinfo{volume}{21}}, \bibinfo{pages}{3969} (\bibinfo{year}{1988}).

\bibitem[{\citenamefont{Zhou and Ernzerhof}(2012)}]{Zhou_Ernzerhof_JPCL_2012}
\bibinfo{author}{\bibfnamefont{Y.}~\bibnamefont{Zhou}} \bibnamefont{and}
  \bibinfo{author}{\bibfnamefont{M.}~\bibnamefont{Ernzerhof}},
  \bibinfo{journal}{J. Phys. Chem. Let.} \textbf{\bibinfo{volume}{3}},
  \bibinfo{pages}{1916} (\bibinfo{year}{2012}).

\bibitem[{\citenamefont{Venkatnathan et~al.}(2000)\citenamefont{Venkatnathan,
  Mishra, and Jensen}}]{Venka_MJ_TCA_2000}
\bibinfo{author}{\bibfnamefont{A.}~\bibnamefont{Venkatnathan}},
  \bibinfo{author}{\bibfnamefont{M.~K.} \bibnamefont{Mishra}},
  \bibnamefont{and} \bibinfo{author}{\bibfnamefont{H.~J.~A.}
  \bibnamefont{Jensen}}, \bibinfo{journal}{Theor. Chem. Acc.}
  \textbf{\bibinfo{volume}{104}}, \bibinfo{pages}{445} (\bibinfo{year}{2000}).

\bibitem[{\citenamefont{Samanta and Yeager}(2008)}]{Samanta_Yeager_JPCB_2008}
\bibinfo{author}{\bibfnamefont{K.}~\bibnamefont{Samanta}} \bibnamefont{and}
  \bibinfo{author}{\bibfnamefont{D.~L.} \bibnamefont{Yeager}},
  \bibinfo{journal}{J. Phys. Chem. B} \textbf{\bibinfo{volume}{112}},
  \bibinfo{pages}{16214} (\bibinfo{year}{2008}).

\bibitem[{\citenamefont{Samanta and Yeager}(2010)}]{Samanta_Yeager_IJQC_2010}
\bibinfo{author}{\bibfnamefont{K.}~\bibnamefont{Samanta}} \bibnamefont{and}
  \bibinfo{author}{\bibfnamefont{D.~L.} \bibnamefont{Yeager}},
  \bibinfo{journal}{Int. J. Quantum Chem.} \textbf{\bibinfo{volume}{110}},
  \bibinfo{pages}{798} (\bibinfo{year}{2010}).

\bibitem[{\citenamefont{Tsednee et~al.}(2015)\citenamefont{Tsednee, Liang, and
  Yeager}}]{Tsednee_LY_PRA_2015}
\bibinfo{author}{\bibfnamefont{T.}~\bibnamefont{Tsednee}},
  \bibinfo{author}{\bibfnamefont{L.}~\bibnamefont{Liang}}, \bibnamefont{and}
  \bibinfo{author}{\bibfnamefont{D.~L.} \bibnamefont{Yeager}},
  \bibinfo{journal}{Phys. Rev. A} \textbf{\bibinfo{volume}{91}},
  \bibinfo{pages}{022506} (\bibinfo{year}{2015}).

\bibitem[{\citenamefont{Zatsarinny et~al.}(2016)\citenamefont{Zatsarinny,
  Bartschat, Fursa, and Bray}}]{Zatsarinny_BFB_JPB_2016}
\bibinfo{author}{\bibfnamefont{O.}~\bibnamefont{Zatsarinny}},
  \bibinfo{author}{\bibfnamefont{K.}~\bibnamefont{Bartschat}},
  \bibinfo{author}{\bibfnamefont{D.~V.} \bibnamefont{Fursa}}, \bibnamefont{and}
  \bibinfo{author}{\bibfnamefont{I.}~\bibnamefont{Bray}}, \bibinfo{journal}{J.
  Phys. B: Atom. Molec. Phys.} \textbf{\bibinfo{volume}{49}},
  \bibinfo{pages}{235701} (\bibinfo{year}{2016}).

\bibitem[{\citenamefont{Knowles et~al.}(1993)\citenamefont{Knowles, Hampel, and
  Werner}}]{Knowles_HW_CC_JCP_1993}
\bibinfo{author}{\bibfnamefont{P.~J.} \bibnamefont{Knowles}},
  \bibinfo{author}{\bibfnamefont{C.}~\bibnamefont{Hampel}}, \bibnamefont{and}
  \bibinfo{author}{\bibfnamefont{H.}~\bibnamefont{Werner}},
  \bibinfo{journal}{J. Chem. Phys.} \textbf{\bibinfo{volume}{99}},
  \bibinfo{pages}{5219} (\bibinfo{year}{1993}).

\bibitem[{\citenamefont{Deegan and Knowles}(1994)}]{Deegan_Knowles_CPL_1994}
\bibinfo{author}{\bibfnamefont{M.~J.} \bibnamefont{Deegan}} \bibnamefont{and}
  \bibinfo{author}{\bibfnamefont{P.~J.} \bibnamefont{Knowles}},
  \bibinfo{journal}{Chem. Phys. Lett.} \textbf{\bibinfo{volume}{227}},
  \bibinfo{pages}{321} (\bibinfo{year}{1994}).

\bibitem[{\citenamefont{Werner et~al.}(2010)\citenamefont{Werner, Knowles,
  Lindh, Knizia, Manby, {Sch\"{u}tz}, and Others}}]{MOLPRO10}
\bibinfo{author}{\bibfnamefont{H.~J.} \bibnamefont{Werner}},
  \bibinfo{author}{\bibfnamefont{P.~J.} \bibnamefont{Knowles}},
  \bibinfo{author}{\bibfnamefont{R.}~\bibnamefont{Lindh}},
  \bibinfo{author}{\bibfnamefont{F.~R.} \bibnamefont{Knizia}},
  \bibinfo{author}{\bibfnamefont{F.~R.} \bibnamefont{Manby}},
  \bibinfo{author}{\bibfnamefont{M.}~\bibnamefont{{Sch\"{u}tz}}},
  \bibnamefont{and} \bibinfo{author}{\bibnamefont{Others}},
  \emph{\bibinfo{title}{{MOLPRO}, version 2010.1, a package of ab initio
  programs}} (\bibinfo{year}{2010}).

\bibitem[{\citenamefont{Prascher et~al.}(2011)\citenamefont{Prascher, Woon,
  Peterson, Dunning, and Wilson}}]{Prascher_WOKDW_TCA_2011}
\bibinfo{author}{\bibfnamefont{B.~P.} \bibnamefont{Prascher}},
  \bibinfo{author}{\bibfnamefont{D.~E.} \bibnamefont{Woon}},
  \bibinfo{author}{\bibfnamefont{K.~A.} \bibnamefont{Peterson}},
  \bibinfo{author}{\bibfnamefont{T.~H.} \bibnamefont{Dunning}},
  \bibnamefont{and} \bibinfo{author}{\bibfnamefont{A.~K.}
  \bibnamefont{Wilson}}, \bibinfo{journal}{Theor. Chem. Acc.}
  \textbf{\bibinfo{volume}{128}}, \bibinfo{pages}{69} (\bibinfo{year}{2011}).

\bibitem[{\citenamefont{Chao et~al.}(1990)\citenamefont{Chao, Falcetta, and
  Jordan}}]{Chao_FY_JCP_1990}
\bibinfo{author}{\bibfnamefont{J.~S.} \bibnamefont{Chao}},
  \bibinfo{author}{\bibfnamefont{M.~F.} \bibnamefont{Falcetta}},
  \bibnamefont{and} \bibinfo{author}{\bibfnamefont{K.~D.}
  \bibnamefont{Jordan}}, \bibinfo{journal}{J. Chem. Phys.}
  \textbf{\bibinfo{volume}{93}}, \bibinfo{pages}{1125} (\bibinfo{year}{1990}).

\bibitem[{\citenamefont{McCurdy et~al.}(1981)\citenamefont{McCurdy, Lauderdale,
  and Mowrey}}]{McCurdy_LM_JCP_1981}
\bibinfo{author}{\bibfnamefont{C.~W.} \bibnamefont{McCurdy}},
  \bibinfo{author}{\bibfnamefont{J.~G.} \bibnamefont{Lauderdale}},
  \bibnamefont{and} \bibinfo{author}{\bibfnamefont{R.~C.}
  \bibnamefont{Mowrey}}, \bibinfo{journal}{J. Chem. Phys.}
  \textbf{\bibinfo{volume}{75}}, \bibinfo{pages}{1835} (\bibinfo{year}{1981}).

\bibitem[{\citenamefont{Gallup}(2011)}]{Gallup_PRA_2011}
\bibinfo{author}{\bibfnamefont{G.~A.} \bibnamefont{Gallup}},
  \bibinfo{journal}{Phys. Rev. A} \textbf{\bibinfo{volume}{84}},
  \bibinfo{pages}{012701} (\bibinfo{year}{2011}).

\bibitem[{\citenamefont{Burrow and Comer}(1975)}]{Burrow_Comer_JPB_1985}
\bibinfo{author}{\bibfnamefont{P.~D.} \bibnamefont{Burrow}} \bibnamefont{and}
  \bibinfo{author}{\bibfnamefont{J.}~\bibnamefont{Comer}}, \bibinfo{journal}{J.
  Phys. B: Atom. Molec. Phys.} \textbf{\bibinfo{volume}{8}},
  \bibinfo{pages}{L92} (\bibinfo{year}{1975}).

\end{thebibliography}

\end{document}